\documentclass[aps,twocolumn,prd,superscriptaddress]{revtex4-1}

\usepackage[utf8]{inputenc}

\usepackage{graphicx}
\usepackage{mathtools}
\usepackage{xcolor}
\usepackage{siunitx}
\usepackage{hyperref}

\def\eq#1{\eqref{#1}}
\def\App#1{App.~\ref{#1}}

\newcommand{\imag}{\text{i}}

\newcommand{\T}{\mathcal{T}} 

\newcommand{\Zc}{Z_c}
\newcommand{\ZA}{Z_A}
\newcommand{\psym}{\bar{p}}

\newcommand{\Acc}{{\bar{c}cA}}
\newcommand{\GammaAcc}{\Gamma^{(3)}_\Acc}
\newcommand{\alphaAcc}{\alpha_\Acc}
\newcommand{\lAcc}{\lambda_\Acc}
\newcommand{\TAccC}{\T_{\Acc}^{\text{cl}}}

\newcommand{\AAA}{{A^3}}
\newcommand{\GammaAAA}{\Gamma^{(3)}_\AAA}
\newcommand{\alphaAAA}{\alpha_\AAA}
\newcommand{\lAAA}{\lambda_\AAA}
\newcommand{\TAAAC}{\T_{\AAA}^{\text{cl}}}

\newcommand{\AAAA}{{A^4}}
\newcommand{\GammaAAAA}{\Gamma^{(4)}_\AAAA}
\newcommand{\alphaAAAA}{\alpha_\AAAA}
\newcommand{\lAAAA}{\lambda_\AAAA}
\newcommand{\TAAAAC}{\T_{\AAAA}^{\text{cl}}}

\def\0#1#2{\frac{#1}{#2}}

\setkeys{Gin}{width=0.48\textwidth}

\newcommand{\eqnewline}{\nonumber\\[1.5ex]}

\graphicspath{{./figures/}}

\newcommand{\gettitle}{Correlation functions of three-dimensional Yang-Mills theory from the FRG}

\hypersetup{
	pdftitle={\gettitle},
	pdfauthor={Corell, Cyrol, Mitter, Pawlowski, Strodthoff},
	pdfkeywords={Yang-Mills theory} 
	{correlations functions} {three dimensions}
	{gluon mass gap} {propagators} 
	{running couplings} {functional renormalisation group},
	bookmarksopen=true,
	bookmarksopenlevel=2,
	bookmarksnumbered=true
}

\begin{document}

\title{\gettitle}

\author{Lukas Corell}
\affiliation{Institut f\"ur Theoretische
  Physik, Universit\"at Heidelberg, Philosophenweg 16, 69120
  Heidelberg, Germany} 
  
\author{Anton K. Cyrol}
\affiliation{Institut f\"ur Theoretische
  Physik, Universit\"at Heidelberg, Philosophenweg 16, 69120
  Heidelberg, Germany} 

\author{Mario Mitter}
\affiliation{Department of Physics, Brookhaven National Laboratory, Upton, NY 11973, United States} 

 \author{Jan M. Pawlowski} 
 \affiliation{Institut f\"ur Theoretische
  Physik, Universit\"at Heidelberg, Philosophenweg 16, 69120
  Heidelberg, Germany} 
 \affiliation{ExtreMe Matter Institute EMMI, GSI, Planckstr. 1,
  D-64291 Darmstadt, Germany}

\author{Nils Strodthoff}
\affiliation{Nuclear Science Division, Lawrence Berkeley 
  National Laboratory, Berkeley, CA 94720, USA}

\begin{abstract}
  We compute correlation functions of three-dimensional Landau-gauge
  Yang-Mills theory with the Functional Renormalisation
  Group. Starting from the classical action as only input, we
  calculate the non-perturbative ghost and gluon propagators as well
  as the momentum-dependent ghost-gluon, three-gluon, and four-gluon
  vertices in a comprehensive truncation scheme.  Compared to the
  physical case of four spacetime dimensions, we need more
  sophisticated truncations due to significant contributions from
  non-classical tensor structures.  In particular, we apply a special
  technique to compute the tadpole diagrams of the propagator
  equations, which captures also all perturbative two-loop effects,
  and compare our correlators with lattice and Dyson-Schwinger
  results.
\end{abstract}

\maketitle

\section{Introduction}
\label{sec:Introduction}
Functional methods such as the Functional Renormalisation Group (FRG)
or Dyson-Schwinger equations (DSEs) are non-perturbative
first-principles approaches to Quantum Chromodynamics (QCD), and they are
complementary to lattice simulations. At finite density the
latter approach is hampered by a sign problem, while the former
approaches face convergence and accuracy problems.  The aim of
the fQCD collaboration~\cite{fQCD:2018-02} is to establish the FRG as a
quantitative continuum approach to QCD, with the phase diagram and the
hadron spectrum as primary applications, see
\cite{Mitter:2014wpa,Braun:2014ata,
  Rennecke:2015eba,Cyrol:2016tym,Cyrol:2017ewj,Cyrol:2017qkl} for
recent works.

Building on the advances made in a previous work in four-dimensional space-time
\cite{Cyrol:2016tym}, we consider Landau-gauge YM theory in three dimensions,
in this work.  Similar to its four-dimensional analogue, it is
asymptotically free and confining.  Upon adding an adjoint scalar, it
corresponds to the dimensionally reduced asymptotic high-temperature
limit of four-dimensional YM theory.  Furthermore, the reduced
dimensionality allows lattice simulations at a considerably reduced
numerical expense, making the three-dimensional theory an interesting
testing case that allows truncation checks in functional approaches.
Therefore, the propagators of three-dimensional YM theory have been
studied intensively on the
lattice~\cite{Cucchieri:1999sz,Cucchieri:2003di,Cucchieri:2004mf,%
Cucchieri:2006tf,Cucchieri:2007rg,Cucchieri:2008qm,Cucchieri:2008fc,%
Maas:2008ri,Cucchieri:2009zt,Maas:2009ph,Maas:2010qw,Maas:2011se,%
Cucchieri:2011ig,Bornyakov:2011fn,Bornyakov:2013ysa,Maas:2014xma,%
Cucchieri:2016qyc},
with
DSEs~\cite{Huber:2007kc,Dudal:2008rm,Alkofer:2008dt,Aguilar:2010zx,%
Aguilar:2013vaa,Cornwall:2015lna,Huber:2016tvc},
and in semi-perturbative
settings~\cite{Tissier:2010ts,Tissier:2011ey,Pelaez:2013cpa}.  Its
vertices have been investigated on the
lattice~\cite{Cucchieri:2006tf,Cucchieri:2008qm} as well as with
continuum methods~\cite{Pelaez:2013cpa,Aguilar:2013vaa,Huber:2016tvc}.

So far, the most advanced results for YM theory in three dimensions
within functional approaches have been obtained in a recent DSE
investigation \cite{Huber:2016tvc}.  There, the coupled system of
equations for the classical tensor structures has been solved
self-consistently.  In terms of the complexity of the truncation, the
investigation \cite{Huber:2016tvc} is comparable to the calculation performed in
\cite{Cyrol:2016tym} for the four-dimensional case, which is more
complicated due to non-trivial renormalisation. The present work
builds on these advances, with a focus on the effects of including
non-classical vertices and tensor structures in the tadpole diagrams
of the gluon and ghost propagator equations.

The paper is organized as follows: In \autoref{sec:setup} we review
the treatment of YM theory with the FRG using a vertex expansion for
the effective action.  We focus on new developments for the inclusion
of the propagator tadpole diagrams.  In \autoref{sec:results} we
discuss our results, which includes a thorough investigation of
apparent convergence and a comparison to DSE and lattice results.  The
conclusion is given in \autoref{sec:summary}.  We check the
independence of the regulator and describe the computational setup in the
appendices.

\section{Yang-Mills Theory from the FRG}
\label{sec:setup}

In this section we review the FRG approach to YM theory using a vertex
expansion for the effective action. Although the overall set-up
follows \cite{Cyrol:2016tym,Cyrol:2017rbo}, we repeat the most important steps for
the convenience of the reader.

\begin{figure}
	\includegraphics[width=.35\textwidth]{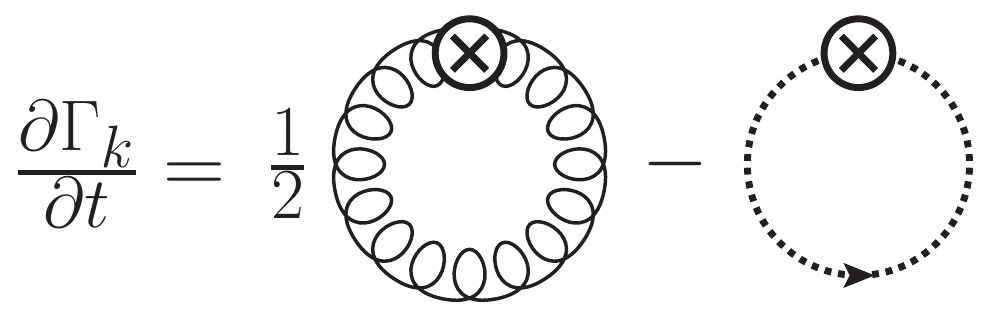}
		\caption{
			Flow equation. Wiggly and dotted lines represent
			the dressed gluon and ghost propagators, respectively.
			The crossed circles denote regulator
			insertions $\partial_t R\,$, see \eq{eq:flow}.}
	\label{fig:FlowEquation}
\end{figure}

The FRG is a non-perturbative continuum method that implements
Wilson's idea of including quantum fluctuations in momentum shells for
the effective action, see
\cite{Berges:2000ew,Pawlowski:2005xe,Gies:2006wv,Schaefer:2006sr,Braun:2011pp}
for QCD-related reviews. The key object in this approach, pioneered by
Wetterich \cite{Wetterich:1992yh}, is the scale-dependent analogue of
the effective action $\Gamma_k$. The RG or infrared cutoff scale $k$
is introducted via a momentum-dependent regulator function $R_k$ that
acts like a fluctuation-suppressing mass term on momentum scales
$p^2\lesssim k^2\,$.  The scale dependence of $\Gamma_k$ is governed
by an exact equation with a simple one-loop structure, 
\begin{align}
	{\partial_t}\Gamma_k[\phi] = 
		\frac{1}{2} \int_p G_{\mu\nu}^{ab}[\phi]\,{\partial_t}R_{\nu\mu}^{ba} 
		-\int_p G^{ab}[\phi]\,{\partial_t}R^{ba}\,,
	\label{eq:flow}
\end{align} 
where $\int_p=\int \text{d}^3p/(2 \pi)^3\,$ and the full field-,
momentum-, and scale-dependent gluon and ghost propagator 
\begin{align}\label{eq:Gprop}
G_k[\phi] =\0{1}{\Gamma^{(2)}[\phi]+R_k}\,,\qquad \Gamma_k^{(n)}[\phi]
=\0{\delta^n\Gamma[\phi]}{\delta\phi^n}\,.
\end{align}
The superfield $\phi=(A_\mu, c,\bar c)$ consists of gauge,
ghost, and anti-ghost fields.  In \eq{eq:flow} the propagators
$G_{ab}^{\mu\nu}[\phi]$ and $G^{ab}[\phi]$ are the diagonal gluon and
off-diagonal ghost--anti-ghost components of the propagator
\eq{eq:Gprop}. A pictorial representation of \eq{eq:flow} is given in
\autoref{fig:FlowEquation}.

\begin{figure*}
	\includegraphics[width=.8\textwidth]{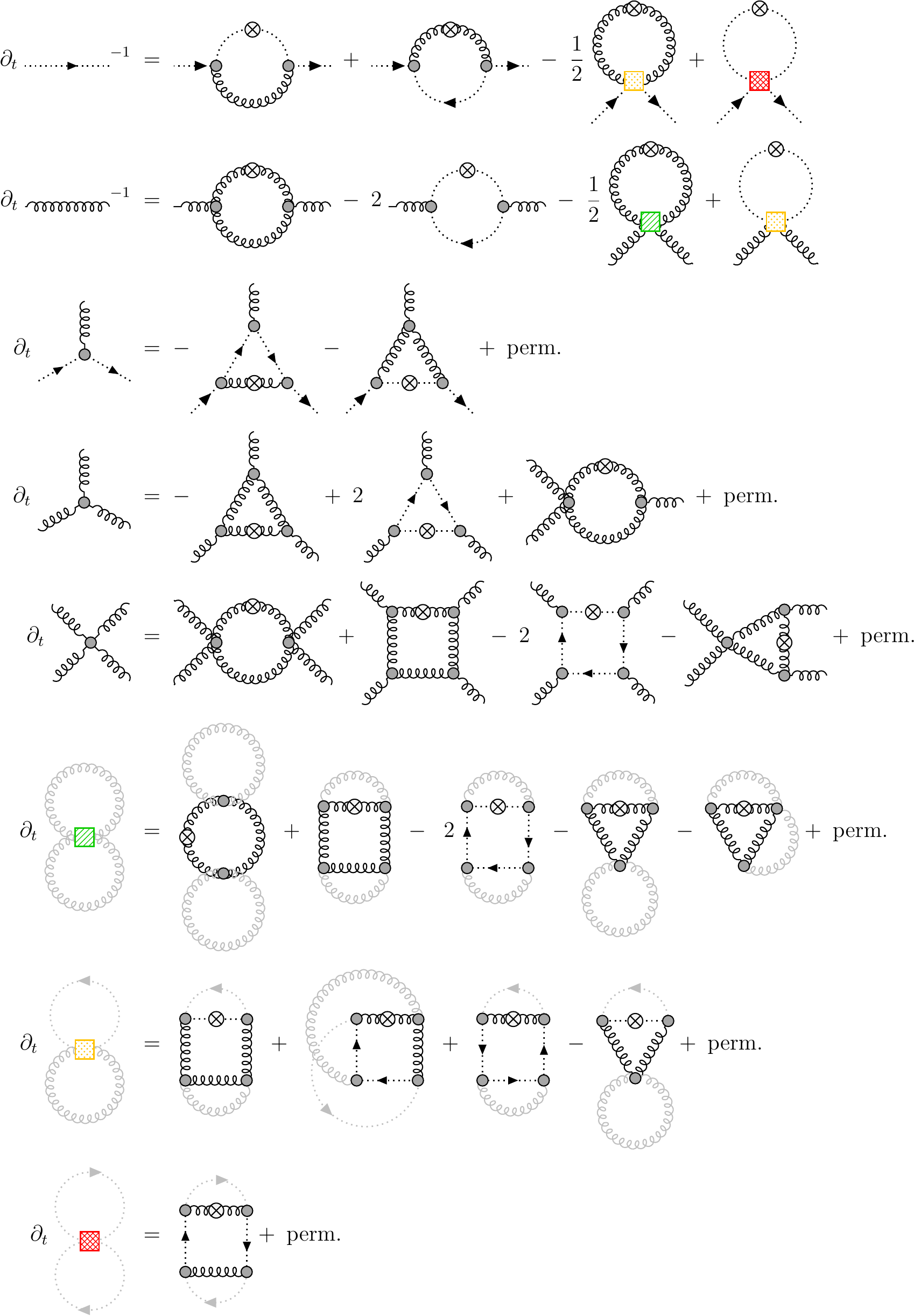}
	\caption{
		Diagrams that contribute to the truncated flows of
		propagators and vertices.  While filled circles denote
		dressed ($1$PI) vertices, the squares denote the tadpole
		vertices explained in \autoref{sec:tadpoleVertices}. Shaded
		lines indicate the projection procedure of the tadpoles
		vertices.  Permutations include not only (anti-)symmetric
		permutations of external legs but also permutations of the
		regulator insertions.
	}
	\label{fig:diagrams}
\end{figure*}

The
regulator functions are given in \App{app:regulator}, where we
also demonstrate the independence of the results from the choice of
the regulator function.  Flow equations for the $1$PI $n$-point functions are
straightforwardly derived from \eq{eq:flow} by taking functional
derivatives with respect to the fields, see \autoref{fig:diagrams} for
the diagrammatic equations.

\subsection{Vertex expansion}
Due to the structure of the flow equation \eq{eq:flow}, the flow
equation for an $n$-point correlator depends on up to $(n+2)-$point
functions.  This leads to an infinite tower of coupled equations,
which have to be truncated within appropriate non-perturbative
expansion schemes in order to be numerically solvable.  As in
\cite{Cyrol:2016tym}, we work in a systematic vertex expansion scheme,
corresponding to an expansion of the effective action in terms of 1PI
correlation functions.  Relying on the structural similarities of the
three-dimensional theory to its four-dimensional analogue, we take all
classical vertices into account, i.e. the ghost-gluon, three- and
four-gluon vertex.  In addition, we compute so-called tadpole vertices
as discussed in the next \autoref{sec:tadpoleVertices}.  For
later reference we quickly recapitulate the parametrisations for the
propagators and classical vertex functions considered in this work.
The gluon and ghost two-point functions 
are parametrised in terms of scalar dressing functions $1/\ZA(p)$ and
$1/\Zc(p)$,
\begin{align}
	[\Gamma^{(2)}_{AA}]^{ab}_{\mu\nu}(p) &= 
		\ZA(p)\, p^2\, \delta^{ab}\, \Pi^{\bot}_{\mu\nu}(p) \,,\eqnewline
	[\Gamma^{(2)}_{\bar c c}]^{ab}(p) &= \Zc(p)\, p^2\, \delta^{ab} \,,
\label{eq:propagators}
\end{align}
where $\Pi^{\bot}_{\mu\nu}(p)=\delta_{\mu\nu}-p_\mu p_\nu/p^2$ denotes
the transverse projection operator. We parametrise the three-point
vertices by
\begin{align}
	[\GammaAcc]^{abc}_\mu (p,q) &= 
		\sqrt{4\pi\,\alpha(\mu)} \, \lAcc(p,\,q) \, [\TAccC]^{abc}_{\mu}(p,q)\,,\eqnewline
	[\GammaAAA]^{abc}_{\mu\nu\rho} (p,q) &= 
		\sqrt{4\pi\,\alpha(\mu)} \, \lAAA(p,\,q) \, [\TAAAC]^{abc}_{\mu\nu\rho}(p,q)\,.
\label{eq:threepoint}
\end{align}
Their classical tensor structures are given by
\begin{align}
	\left[\TAccC\right]^{abc}_{\mu}(p,q) &=
		\imag f^{abc} q_\mu  \,, \eqnewline
	\left[\TAAAC\right]^{abc}_{\mu\nu\rho}(p,q) &= 
		\imag f^{abc} \left\{(p-q)_\rho \delta_{\mu\nu}+ \text{ perm.}\right\} \,.
\end{align}
The transversely projected basis for the ghost-gluon vertex consists of only
one single element, whereas the corresponding basis for the three-gluon vertex counts four
elements. The impact of non-classical tensor structures in the three-gluon vertex
have been found to be subleading \cite{Eichmann:2014xya} in four space-time
dimensions. Here we assume that they are also subleading in three dimensions and
neglect them.
The parametrisation of the four-gluon vertex is given by
\begin{align}
	[\GammaAAAA]^{abcd}_{\mu\nu\rho\sigma}(p,q,r) &= 
		4\pi\,\alpha(\mu) \, \lAAAA (\bar{p})\, [\TAAAAC]^{abcd}_{\mu\nu\rho\sigma}\,,
\label{eq:fourgluon}
\end{align}
where the classical tensor structure is given by
\begin{align}
	\left[\TAAAAC\right]^{abcd}_{\mu\nu\rho\sigma} &=
		f^{abn}f^{cdn}\delta_{\mu\rho}\delta_{\nu\sigma} + \text{perm.}\,.
\end{align}
The inclusion of non-classical tensor structures in the four-gluon vertex 
is discussed below in \autoref{sec:tadpoleVertices}. The four-gluon dressing function(s) are approximated as a 
function of the average momentum $\bar p^2=\frac{1}{4}(p_1^2+p_2^2+p_3^2+p_4^2)$ which was shown
to be a good approximation for the full momentum dependence in 
four space-time dimensions \cite{Cyrol:2014kca}
and we assume that the same holds in three dimensions.

\begin{figure*}
  \includegraphics{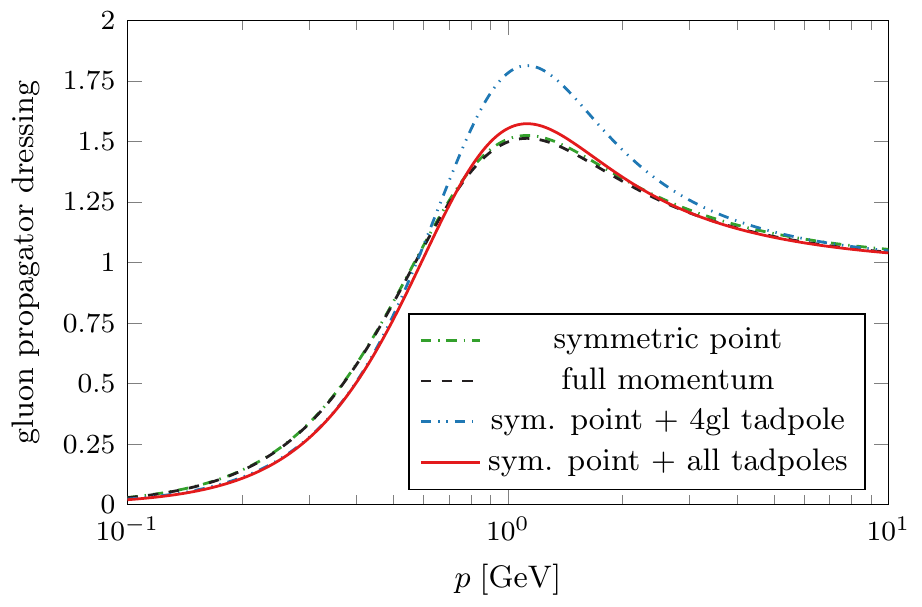}
  \hfill
  \includegraphics{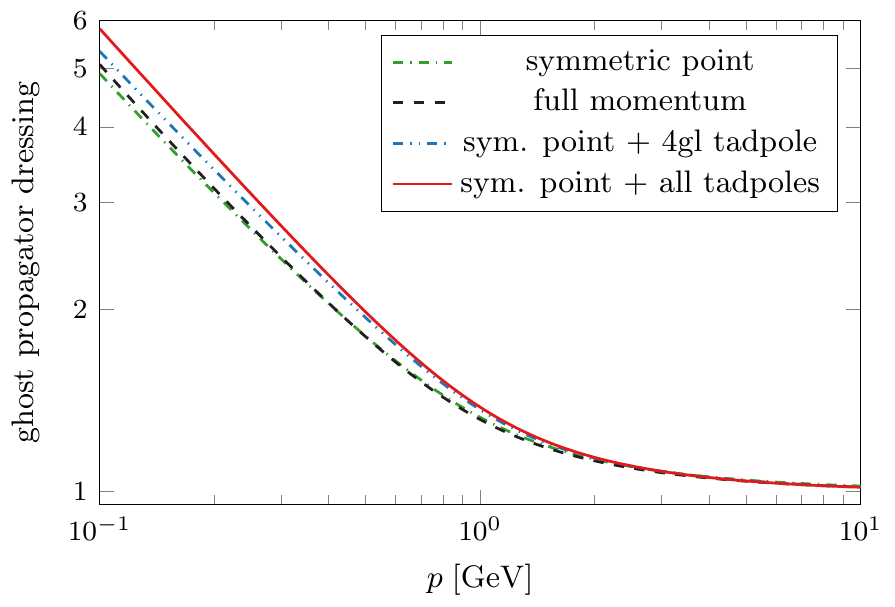}
  \caption{Truncation dependence of the gluon propagator
    dressing $1/\ZA(p)$ (left) and ghost propagator dressing $1/\Zc(p)$ (right).
    \emph{Symmetric~point} and \emph{full momentum} denotes using the average
    momentum and full momentum dependency, respectively, in the
    three-gluon vertex. Results with \emph{+~4gl~tadpole} and \emph{+~all tadpoles} include the respective tadpole
    diagrams.}
  \label{fig:truncation_comparison}
\end{figure*}

From the momentum-dependent dressing functions of the different
correlators, we can define corresponding running couplings via
\begin{align}
	\label{eq:running_couplings}
	\alphaAcc(p) &= \alpha(\mu) \frac{\lAcc^2(p)}{\ZA(p)\,\Zc^2(p)}\,,\eqnewline
	\alphaAAA(p) &= \alpha(\mu) \frac{\lAAA^2(p)}{\ZA^3(p)}\,,\eqnewline
	\alphaAAAA(p) &= \alpha(\mu) \frac{\lAAAA(p)}{\ZA^2(p)}\,.
\end{align}
Due to gauge invariance, encoded in the Slavnov-Taylor identities, all the couplings 
\eq{eq:running_couplings} have to agree in the perturbative regime of the theory.  
Furthermore, the dimensional
suppression of the running coupling ensures that the dressing
functions take their bare values at large momentum scales,
\begin{align}\label{eq:UV-vert}
	\lim_{p\to\infty} \lAcc(p) = \lim_{p\to\infty} \lAAA(p) = \lim_{p\to\infty} \lAAAA(p) = 1\,, 
\end{align}
for UV-trivial wave function renormalisations 
\begin{align}\label{eq:UV-wave}
\lim_{p\to\infty}\ZA(p)\to 1\,,\qquad \lim_{p\to\infty}\Zc(p)\to 1\,.
\end{align}

The truncation described above depends only trivially on the gauge group.
In particular, only the quadratic casimir of the adjoint representation appears
in the truncated set of equations. Therefore, it can be absorbed into a redefinition 
of the coupling, which in turn can be turned into a redefinition of the physical scale, 
see \cite{Cyrol:2017qkl,Cyrol:2017rbo} for a more detailed discussion.
The same holds for the extended truncation described in the next subsection.
Thus, our results are effectively independent of the gauge group.
However, this does not indicate a bad truncation since also in perturbation theory YM theory
is independent of the gauge group up to three loops, see e.g. \cite{Herzog:2017ohr}
for a recent discussion. Also the DSE results from \cite{Huber:2016tvc} do not possess a genuine 
gauge group dependence and lattice results for the propagators show only a mild dependence
on the gauge group \cite{Cucchieri:2007zm,Sternbeck:2007ug}. Consequently, we compare our results 
to $SU(2)$ lattice results.

\subsection{Tadpole vertices}
\label{sec:tadpoleVertices}

The structure of the flow equation \eq{eq:flow} implies that fully
dressed four-point functions appear on the right hand side of the
propagator equations, see \autoref{fig:diagrams}.  In general, this
requires the full knowledge of all momentum-dependent non-classical
four-point tensor dressings.  Although some exploratory studies
exist~\cite{Kellermann:2008iw,Binosi:2014kka,Gracey:2014ola,Cyrol:2014kca,Huber:2017txg},
their dynamical back-coupling into the propagator equations has still
not been achieved.  In the following, we propose a method that
captures most of the dynamics on the level of the propagator
equations, while it keeps the numerical effort at a manageable level.
As an example, we consider the gluon tadpole contribution to the gluon
propagator equation.  All other tadpole diagrams are obtained
analogously.  The gluon tadpole contribution to the flow of the gluon
two-point function is given by
\begin{align}
	\label{eq:tadpoleStart}
	\partial_t [\Gamma_{A^2}^{(2)}]^{ab}_{\mu\nu}(p) = 
		\frac{1}{2} \int_p  \;
		[\GammaAAAA]^{abcd}_{\mu\nu\rho\sigma}(p,-p,q) \cdot
		[G\,\partial_t R\, G]^{dc}_{\sigma\rho}(q)\,.
\end{align}
Exploiting that the gluon propagator is diagonal in colour space and
transverse with respect to its momentum in Landau gauge, we can project
\eq{eq:tadpoleStart} with $\delta^{ab}\,\Pi^{\bot}_{\mu\nu}(p)\,$.
From this we see that the gluon propagator equation depends only on
the projected four-point function
\begin{align}
	\label{eq:tadpoleKey}
	T_{A^4}(p,q) &= \Pi^{\bot}_{\mu\nu}(p)
	[\GammaAAAA]^{abcd}_{\mu\nu\rho\sigma}(p,-p,q)\,\Pi^{\bot}_{\rho\sigma}(q)\,.
\end{align}
Therefore, the full contribution of the four-gluon vertex to the
tadpole is already contained in this single scalar function, whose
flow we can compute directly from projecting the corresponding
equation accordingly, cf. \autoref{fig:diagrams}.  In particular, this
procedure includes the back-coupling effect of all non-classical
tensor structures that are generated at the perturbative one-loop
level, including therefore also all two-loop effects of the tadpole
diagrams in the propagator equations.  The non-classical tensor
structures couple back into the vertices indirectly via the
propagators.  We neglect their direct back-coupling into the vertex
equations.  However, we expect this approximate treatment to yield a
considerable improvement of the truncation at comparably moderate
numerical costs.

\begin{figure*}
	\includegraphics{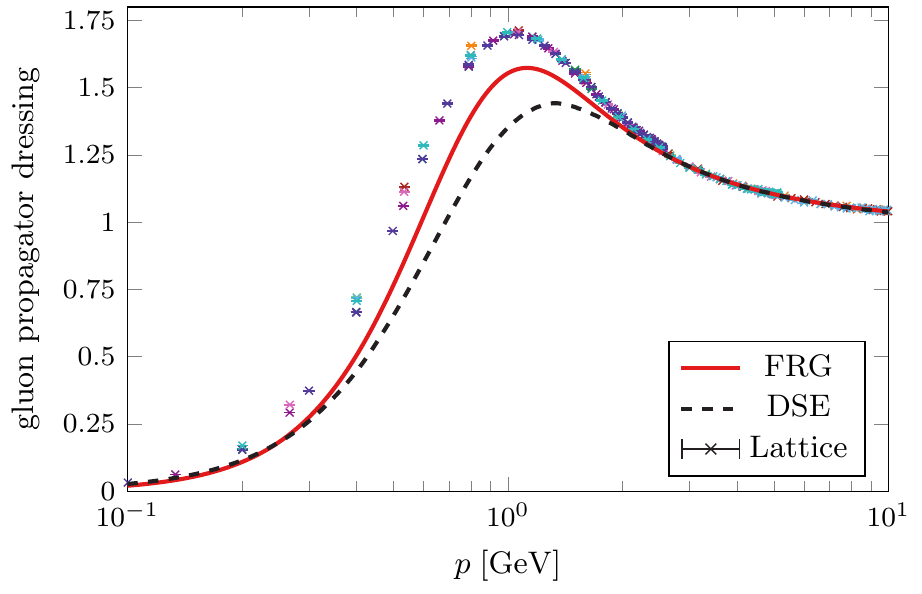}
	\hfill
	\includegraphics{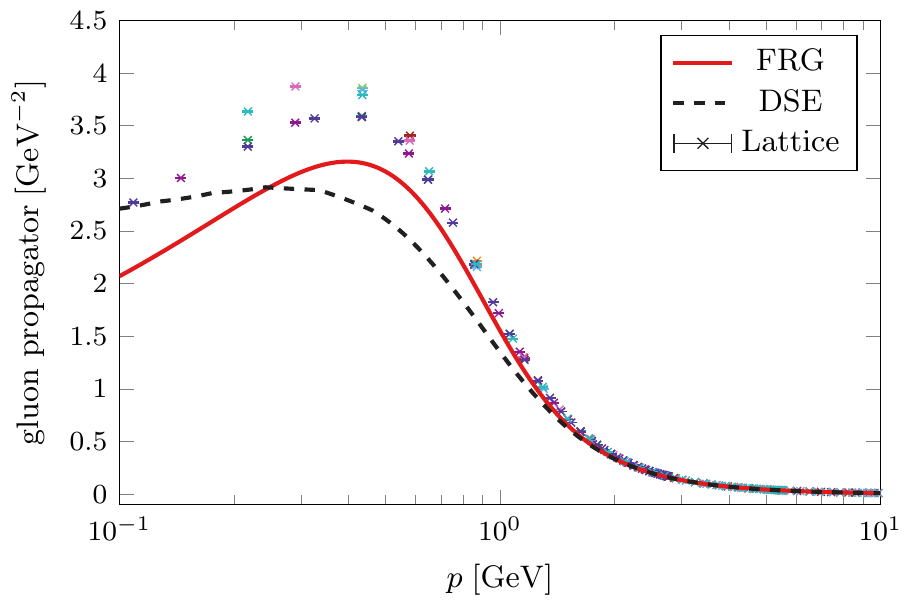}
	\caption{
		Gluon propagator dressing $1/\ZA(p)$ (left) and the dimensionful propagator $1/(p^2\ZA(p))$ (right) in comparison with DSE \cite{Huber:2016tvc} and lattice \cite{Cucchieri:2006tf,Cucchieri:2008qm,Maas:2014xma,Maas:preparation} results.
	}
	\label{fig:Gluon}
\end{figure*}

\section{Results}
\label{sec:results}

In this section we present the main findings of our
investigation. Our solutions are of the scaling type, 
and are obtained as described in \App{app:workflowandsolution}. 
After discussing the truncation dependence of our
results we provide an extensive comparison to results from lattice
gauge theory and Dyson-Schwinger equations.  We close with a
determination of the infrared scaling coefficients and their
comparison to those of finite temperature Yang-Mills theory in four
dimensions.

\subsection{Truncation and Apparent Convergence}
\label{sec:results:trunc}

In order to assess the influence of the truncation on our results, we compare
three different extensions of our simplest symmetric point approximation:
\begin{enumerate}
	\item \emph{symmetric point}: only classical vertices with dressing functions
		that depend only on the symmetric momentum configuration,
	\item \emph{full momentum}: same as 1. symmetric point, but including the full
		momentum dependence of the ghost-gluon and three-gluon vertex dressings,
	\item \emph{sym.~point~+~4gl tadp.}: same as 1., but with the effects of the
		non-classical tensors of the four-gluon-vertex included in the
		tadpole diagram of the gluon propagator equation
		as described in \autoref{sec:tadpoleVertices},
	\item \emph{sym.~point~+~all tadp.}: same as 3., but additionally including 
		the effects of the two-ghost-two-gluon and four-ghost vertices
		in both propagator equations, 
		see \autoref{sec:tadpoleVertices} and \autoref{fig:diagrams} for a visualization.
\end{enumerate}
The corresponding results for the propagators are shown in
\autoref{fig:truncation_comparison}. The first immediate observation
is that the additional momentum dependence (2.) in the three-gluon and
ghost-gluon vertices does not visibly affect the propagators. On the
contrary, the full momentum dependence and tensor structures of the
four-point functions in the tadpole diagrams significantly affect the
propagators.  Concerning the goal of apparent convergence, we observe
that including the tadpole contribution of the four-gluon vertex alone
has a comparably pronounced effect, most of which is counteracted by
the remaining tadpoles.  This indicates that a fast convergence may be
achieved if the underlying consistent resummation pattern is preserved 
within the truncation scheme. A similar observation
has already been made in the matter sector of QCD in four space-time
dimensions \cite{Mitter:2014wpa,Cyrol:2017ewj}. There, it is found
that the effect of non-classical tensor structures in the quark-gluon
vertex is counter-acted by corresponding structures in higher
quark-gluon interactions that stem from the same BRST-invariant
operator. We conclude that it is of chief importance to fully reveal these 
resummation patterns.

\begin{figure*}
	\includegraphics{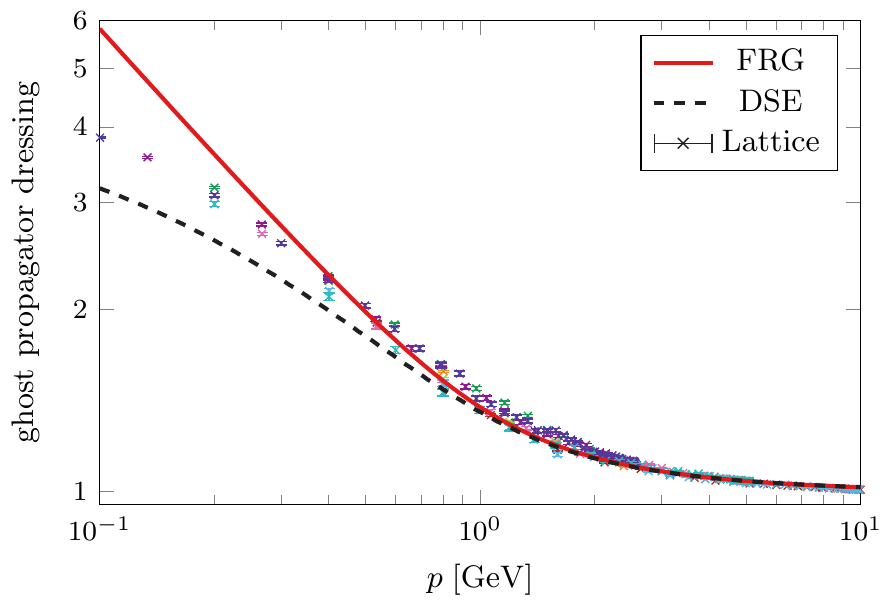}
	\hfill
	\includegraphics{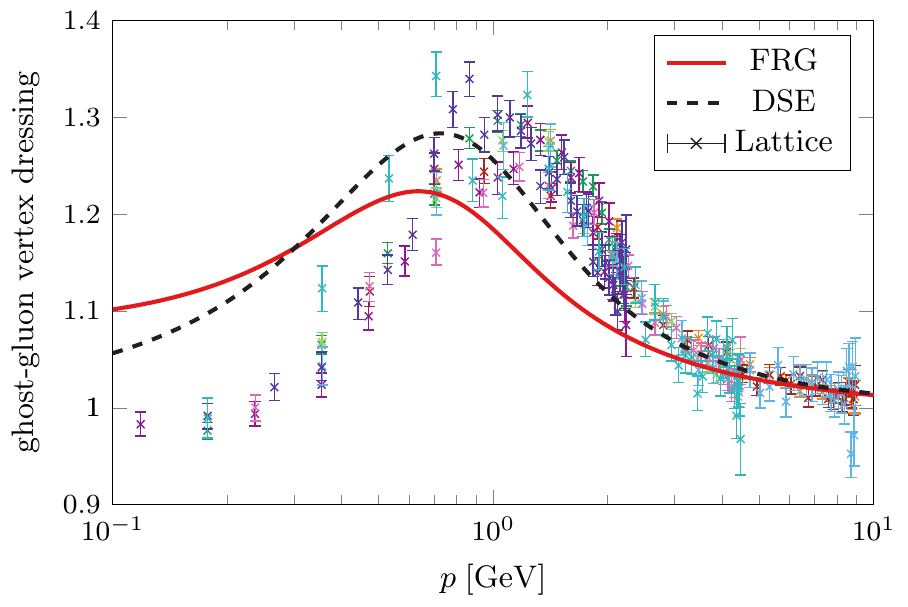}
	\caption{
		Ghost propagator dressing $1/\Zc(p)$ (left) and ghost-gluon 
		vertex dressing $\lAcc(\psym)$ (right) compared to 
		DSE \cite{Huber:2016tvc} and lattice 
		\cite{Cucchieri:2006tf,Cucchieri:2008qm,Maas:2014xma,Maas:preparation} results.
	}
	\label{fig:Ghost}
\end{figure*}

\subsection{Comparison to DSE and Lattice}
\label{sec:results:comp}

In this section we compare the results from our most extensive
truncation, 4.~\emph{sym.~point~+~all~tadp.} (see
\autoref{sec:results:trunc}), to results obtained from $SU(2)$ lattice
gauge theory
\cite{Cucchieri:2006tf,Cucchieri:2008qm,Maas:2014xma,Maas:preparation}
and with Dyson-Schwinger equations \cite{Huber:2016tvc}.  To that end
we normalise both, lattice and DSE results respective to our results
in the UV regime, for more details see \App{app:fits}. We emphasise
again hat the presented FRG result is of the scaling type
\cite{vonSmekal:1997ohs,Zwanziger:2001kw,Lerche:2002ep,Fischer:2002eq,Pawlowski:2003hq,Alkofer:2004it,Fischer:2006vf,Alkofer:2008jy,Fischer:2009tn},
whereas the lattice and DSE results are decouplings solutions
\cite{Cucchieri:2007rg,Cucchieri:2008fc,Aguilar:2008xm,Boucaud:2008ky,Maas:2009ph},
characterised by a finite, non-vanishing value of the gluon propagator
at $p=0\,$.

\subsubsection{Propagators}
\label{sec:YM3d:Results:Propagators}

From \autoref{fig:Gluon} and the left panel of \autoref{fig:Ghost},
it is clearly seen that our results agree well with the rescaled lattice results in
the UV regime with a discrepancy arising below $\SI{3}{\GeV}$. This difference is most
likely due to truncation artifacts in our results which has to be clarified in future work.
The most obvious culprit are missing effects in the equations for the classical 
vertex tensor structures due to the leading non-classical tensor structures
of the three- and four-point functions.

The DSE gluon propagator from \cite{Huber:2016tvc} has a smaller bump
than both the FRG and lattice propagators. In
\autoref{sec:results:trunc} we have shown that non-classical tensor
structures have the net effect of increasing the bump in the gluon
propagator. In comparison to the DSE truncation in
\cite{Huber:2016tvc}, the present approximation includes more
non-classical tensor structures. Although this may serve as an explanation,
the system of equations is highly non-linear, and such an incomplete comparison 
is potentially misleading. Another
factor may be that the DSE results are of the decoupling type whereas
our results are of the scaling type, which generically
show a larger bump~\cite{Cyrol:2016tym}.  In order to
perform a more informative comparison between the DSE and FRG results, a DSE scaling solution would
be preferable because of its uniqueness 
\cite{Fischer:2006vf,Fischer:2009tn}.

\subsubsection{Vertices}

\begin{figure*}
	\includegraphics{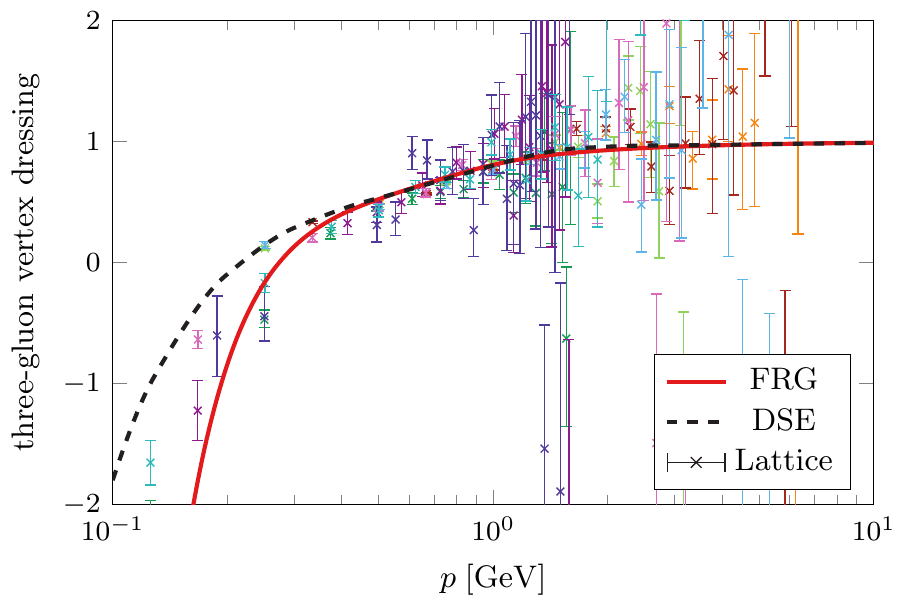}
	\hfill
	\includegraphics{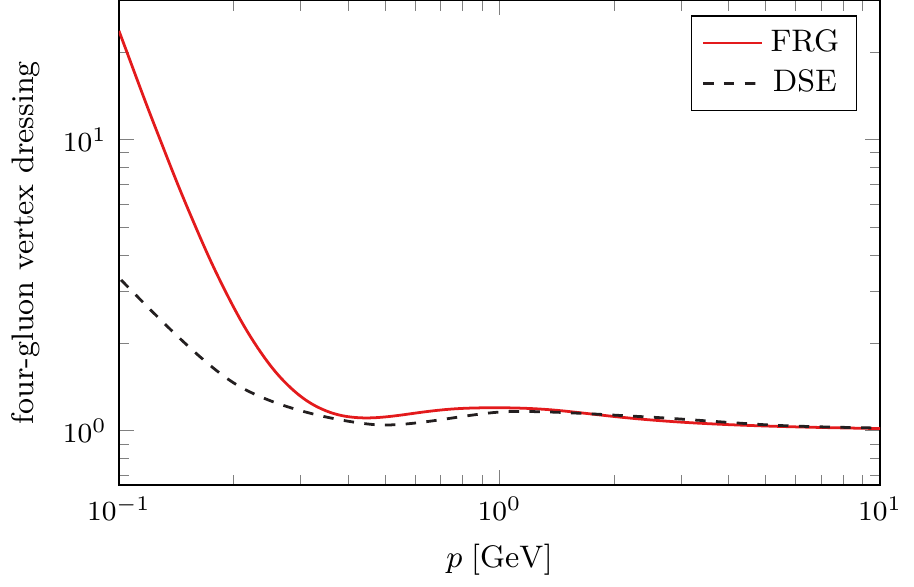}
	\caption{
		Three-gluon (left) and four-gluon (right) vertex dressings, $\lAAA(\psym)$ and $\lAAAA(\psym)\,$, compared to DSE \cite{Huber:2016tvc} and lattice \cite{Cucchieri:2006tf,Cucchieri:2008qm,Maas:2014xma,Maas:preparation} results.
	}
	\label{fig:GluonicVertices}
\end{figure*}

The ghost-gluon and gluonic vertex dressings are shown in comparison with DSE~\cite{Huber:2016tvc} and lattice~\cite{Cucchieri:2006tf,Cucchieri:2008qm,Maas:2014xma,Maas:preparation} 
results in \autoref{fig:Ghost} and \autoref{fig:GluonicVertices}, where the momentum scale was
set using the fit parameters from the gluon propagator in the last section.
Similar to the propagators, all dressings converge to unity in the ultraviolet.

Concerning the ghost-gluon vertex dressing, we find that the lattice result has its peak at a higher 
scale than the dressings computed with functional methods. A similar but, at least in the FRG result less
obvious, deviation can be observed already in the ghost propagator dressing, indicating a general scale
mismatch between ghost- and glue sector. This is particularly interesting, since also recent QCD investigations
with very sophisticated truncation schemes \cite{Mitter:2014wpa,Cyrol:2017ewj}
show such a scale mismatch between the matter sector and the glue part of the theory, whereas the glue sector in itself 
runs consistently. We think that in both cases, missing higher-order effects are the most likely source of these deviations.

The FRG three-gluon vertex dressing shows very good agreement with the lattice results over all momenta.
In particular, the agreement in the infrared is surprising, since the lattice features a decoupling solution, 
which has a linearly divergent three-gluon vertex dressing function~\cite{Pelaez:2013cpa,Aguilar:2013vaa,Huber:2016tvc}, 
whereas our solution is the scaling solution, which has a stronger divergence in the infrared,
$\lAAA(p)\propto\left(p^2\right)^{-3\kappa-1/2}\,$, cf. \autoref{sec:YM3d:Results:Scaling}.
The FRG and DSE four-gluon vertices agree well, whereas lattice measurements of the four-gluon vertex are not available as of now.

\subsection{Infrared Scaling Exponents}
\label{sec:YM3d:Results:Scaling}

In the scaling solution, all correlators scale with a specific power law in the infrared.
It can be shown that self-consistency demands that the anomalous scaling behavior of any $(2n+m)$-point 
function with $2n$ ghost and $m$ gluon legs in $d$ dimensions is determined by one single
scaling exponent and can be written as~\cite{Fischer:2006vf,Huber:2007kc,Fischer:2009tn}
\begin{align}
	\label{eq:GeneralScaling}
	\lim_{p\rightarrow 0} \; \lambda^{(2n,m)}(p)
		\propto \left(p^2\right)^{(n-m)\kappa+(1-n)\left(\frac{d}{2}-2\right)}\,.
\end{align}
In particular, for the two-point functions, the scaling power laws are then given by~\cite{Zwanziger:2001kw,Lerche:2002ep}
\begin{align}
	\label{eq:GeneralPropScaling}
	&\Gamma_{\bar c c}(p) \propto p^2 \cdot \left(p^2\right)^{\kappa}\eqnewline
	&\Gamma_{AA}(p) \propto p^2 \cdot \left(p^2\right)^{-2\kappa+\frac{d}{2}-2}\,,
\end{align}
where we took their canonical scaling into account.
The right panel of \autoref{fig:Gluon} and the left panel of
\autoref{fig:Ghost} clearly reveal the
power law behaviour.  Fitting the propagators with
\eq{eq:GeneralPropScaling}, we obtain the three-dimensional scaling
exponents,
\begin{align}
	\label{eq:kappa3d}
	\kappa_\text{\tiny sym.~p.}=0.321 \pm 0.001\,,\eqnewline
	\kappa_\text{\tiny full~mom.}=0.348 \pm 0.013\,,\eqnewline
	\kappa_\text{\tiny sym.~p.~+~tad.}=0.349 \pm 0.003\,,
\end{align}
for the different truncations. The uncertainty stems from the
difference of the ghost and gluon propagator fits.  In contrast to the
large- and mid-momentum behaviour of the correlators, the scaling
coefficient is also susceptible to the full momentum dependence of the
vertices.

We also compare these scaling coefficients with those of
four-dimensional Yang-Mills theory at finite temperature
\cite{Cyrol:2017qkl}. There an approximation similar to the
symmetric point approximation, (1) in
\autoref{sec:results:trunc}, was used. Fitting the magnetic part of gluon
propagators to the scaling formula \eq{eq:GeneralPropScaling} yields
$\kappa_T = 0.323(3)$. Hence, the magnetic scaling exponent agrees
very well with the scaling exponent of the three-dimensional
theory in the approximation (1). This is expected from
dimensional reduction, and yields a very consistent picture.

\section{Conclusion}
\label{sec:summary}

We have presented non-perturbative correlators of three-dimensional
Landau-gauge Yang-Mills theory obtained from first principles with the
functional renormalisation group. We have checked the reliability of
the results by comparing to lattice results and achieved better
agreement by including non-classical tensors structures in the
truncation scheme. However, at lower momenta the functional and the
lattice results still show a discrepancy of \SI{10}{\percent}. This
hints at sizeable truncation artifacts in three dimensional Yang-Mills 
theory with functional methods at the current truncation level.

These findings are particularly interesting, because an analogous
investigation with the FRG in four dimensions shows considerably
better agreement with the corresponding lattice results already at a
simpler truncation level, based on classical tensor structures
only. This indicates that apparent convergence is achieved with less
effort in the four-dimensional theory. A possible explanation are the
stronger infrared effects that are generically present in lower
dimensions. Phrased differently, the three-dimensional theory 
features a weakened RG irrelevance of the operators corresponding to the
non-classical vertex components.

Interestingly, the effects of non-classical tensors seem to cancel
largely. Although individual contributions result in large
corrections, their overall effect is relatively small but notable. In
this work this is explicitly shown in the propagator tadpole
contributions, whose overall effect is small, when compared to the
individual contributions.  A similar observation has also been made in
the matter sector of four-dimensional QCD for the effect of
non-classical quark-gluon interactions
\cite{Mitter:2014wpa,Cyrol:2017ewj}.  This finding is particularly
important for devising quickly converging truncation schemes
by preserving the underlying resummation patterns.

\acknowledgments

We thank Markus Q. Huber and Axel Maas for discussions.  This work is
supported by ExtreMe Matter Institute (EMMI), the Austrian Science Fund (FWF) through
Erwin-Schr\"odinger-Stipendium No.\ J3507-N27, the Studienstiftung des
deutschen Volkes, the German Research Foundation (DFG) through grants STR 1462/1-1 and MI 2240/1-1,
the U.S. Department of Energy under contract de-sc0012704, and in part
by the Office of Nuclear Physics in the US Department of Energy's
Office of Science under Contract No. DE-AC02-05CH11231. It is part of
and supported by the DFG Collaborative Research Centre "SFB 1225
(ISOQUANT)".

\appendix

\section{Regulator independence}
\label{app:regulator}

\begin{figure}
	\includegraphics{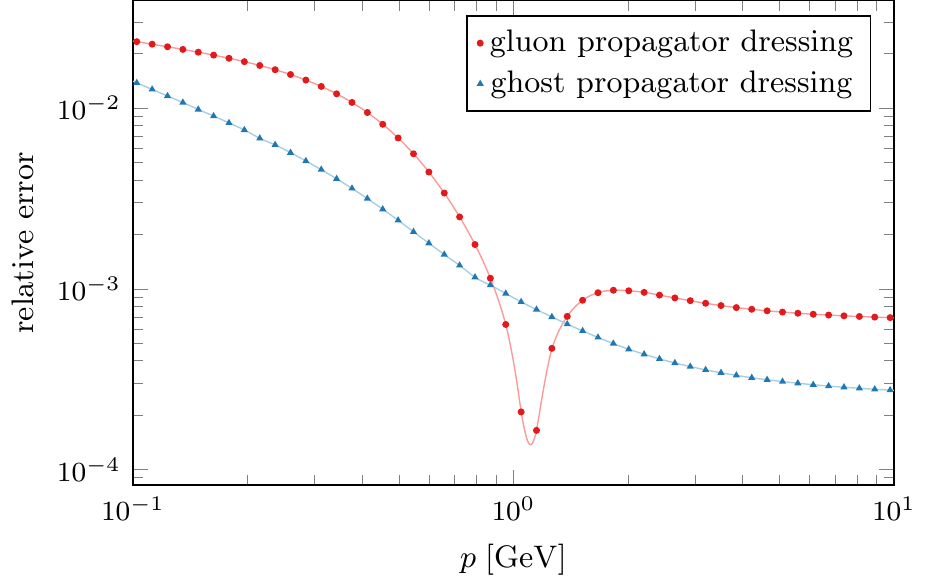}
	\caption{
		Relative errors $\Delta_\mathrm{rel}$ of propagator dressings obtained with different regulator shape functions, given in \eq{eq:exp_regulator_shape_function} and \eq{eq:flat_regulator_shape_function}, in the symmetric point approximation.
	}
	\label{fig:Appendix}
\end{figure}

To check the stability of our results,
we repeat the computations above with the flat \cite{Litim:2000ci} instead of the exponential
regulator shape function.  We parametrise the ghost and gluon
regulators by
\begin{align}
	&R^{ab}(p) = p^2\, \delta^{ab}\, r\left(\frac{p^2}{k^2}\right)\,,\eqnewline
	&R_{\mu\nu}^{ab}(p) = p^2\, \delta^{ab}\,\Pi^{\bot}_{\mu\nu}\, r\left(\frac{p^2}{k^2}\right)\,.
\end{align}
The exponential shape function is given by
\begin{align}
	r_\mathrm{exp}(x)=\frac{x^{m-1}}{\exp\left(x^m\right)-1}\,,
	\label{eq:exp_regulator_shape_function}
\end{align}
whereas the flat one is given by
\begin{align}
	r_\mathrm{flat}(x)=\left(x^{-1}-1\right) \cdot \theta\left(x^{-1}-1\right)\,.
	\label{eq:flat_regulator_shape_function}
\end{align}
The dependence of propagator dressings on the regulator shape functions
is shown in \autoref{fig:Appendix} as relative errors, defined
by
\begin{align}
  \Delta_\mathrm{rel}^2=2\cdot\frac{(\mathcal{O}_\mathrm{exp} - 
  \mathcal{O}_\mathrm{flat})^2}{\mathcal{O}_\mathrm{exp}^2 + \mathcal{O}_\mathrm{flat}^2}\,.
\end{align}
Clearly, the relative errors are well below the percent
level in the IR, and even smaller in the mid-momentum and UV
regimes that are relevant for hadronic observables. Importantly, the
regulator dependence is significantly smaller than the truncation
dependence.

Explicitly demonstrating regulator independence is a standard quality and self-consistency check for truncations 
in the FRG. It is a necessary but not sufficient criterion for the convergence 
of a given truncation. Indeed, we observe that the dependence of our results 
on the regulator shape function is negligible although the truncations are 
not yet converged. Nonetheless, this regulator independence already at low truncation 
orders is a very welcome property.

\section{Numerical computation}
\label{app:workflowandsolution}
Landau gauge has the convenient property that the transverse
correlation functions close among themselves
\cite{Fischer:2008uz,Cyrol:2016tym}, i.e.  correlators with at least
one longitudinal leg do not couple back into the transverse subsystem.  In the
presence of a regulator term, the BRST symmetry is encoded in modified
Slavnov-Taylor identities.  Their most important consequence is a
non-vanishing gluon mass term at finite cutoff scales
\cite{Ellwanger:1994iz}.  Here we present only results for one choice
of the gluon mass term, determined uniquely by the scaling solution
\cite{Zwanziger:2001kw,Lerche:2002ep}.  The consequences of other
choices for the gluon mass term are qualitatively similar to
YM theory in four space-time dimensions and we refer to the
discussion presented in \cite{Cyrol:2016tym} for details.

This work relies on the workflow established within the fQCD collaboration \cite{fQCD:2018-02},
see \cite{Cyrol:2016tym} for details.
Symbolic flow equations were derived using \textit{DoFun}~\cite{Huber:2011qr},
traced using \textit{FormTracer}~\cite{Cyrol:2016zqb}, 
which makes use of FORM~\cite{Vermaseren:2000nd} and its optimization procedure \cite{Kuipers:2013pba}.

\section{Scale setting and normalisation}\label{app:fits}

For comparison, the DSE and lattice results for the propagators in
\autoref{sec:results} are normalised in amplitude and momentum scale
relative to the FRG results. To that end we normalise the DSE/lattice gluon dressings with a least squares
fit to the FRG gluon propagator dressing in
the range \SI{3}{\GeV} to \SI{6}{GeV} with 
\begin{align}
\label{eq:gluonFit}
\min_{c_A,\,c_p}\Big\lbrace\sum_{p_{i,\mathrm{lattice}}}\big(c_\mathrm{A}
\,Z_{A,\,\mathrm{FRG}}^{-1}(c_\mathrm{p}\,p_i)-Z_{A,\,\mathrm{lat/DSE}}^{-1}(p_i)
\big)^2\Big\rbrace\,.
\end{align}
Here, $c_A$ normalises the amplitude while $c_p$ normalises the momentum scale. 
The momentum scale normalisation has to be used for all correlation functions. Hence it is only left to 
fix the amplitudes for the other correlation functions. In particular 
the amplitude of the ghost propagator dressing is normalised with 
\begin{align}
\label{eq:ghostFit}
\min_{c_c}\Big\lbrace\sum_{p_{i,\mathrm{lattice}}}\big(c_\mathrm{c}\,
Z_{c,\,\mathrm{FRG}}^{-1}(c_p\,p_i)-Z_{c,\,\mathrm{lat/DSE}}^{-1}(p_i)\big)^2\Big\rbrace\,.
\end{align}
The lattice results for the vertices have large
statistical lattice error, and we refrain from normalising the 
amplitudes. The dressing of the DSE vertices is trivial for large momenta. 

\newpage

\bibliography{../bib_master}

\end{document}